\documentclass[nonacm, sigconf]{acmart}

\usepackage[english]{babel}
\usepackage[utf8]{inputenc}
\usepackage{tikz-cd}
\usepackage{lipsum}
\usepackage{wasysym}
\usepackage{multirow}
\usepackage{tcolorbox}

\usepackage{hyperref}
\usepackage{array}                  
\usepackage{verbatim}
\newcolumntype{L}[1]{>{\raggedright\let\newline\\\arraybackslash\hspace{0pt}}m{#1}}
\newcolumntype{C}[1]{>{\centering\let\newline\\\arraybackslash\hspace{0pt}}m{#1}}
\newcolumntype{R}[1]{>{\raggedleft\let\newline\\\arraybackslash\hspace{0pt}}m{#1}}
\newcolumntype{J}[1]{>{\let\newline\\\arraybackslash\hspace{0pt}}m{#1}}

\AtBeginDocument{%
  \providecommand\BibTeX{{%
    \normalfont B\kern-0.5em{\scshape i\kern-0.25em b}\kern-0.8em\TeX}}}

\begin{document}



\title{Uncovering the Hidden Potential of Event-Driven Architecture: \\ A Research Agenda}

\author{Luan Lazzari}
\email{luanlazzari@edu.unisinos.br}
\affiliation{%
  \institution{Universidade do Vale do Rio dos Sinos}
  \city{São Leopoldo}
  \state{Rio Grande do Sul}
  \country{Brazil}
}

\author{Kleinner Farias}
\email{kleinnerfarias@unisinos.br}
\affiliation{%
  \institution{Universidade do Vale do Rio dos Sinos}
  \city{São Leopoldo}
  \state{Rio Grande do Sul}
  \country{Brazil}
}

\begin{abstract}
Event-driven architecture has been widely adopted in the software industry, emerging as an alternative to modular development to support rapid adaptations of constantly evolving systems. However, little is known about the effects of event-driven architecture on performance, stability, and software monitoring, among others. Consequently, professionals end up adopting it without any empirical evidence about its impact. Even worse, the current literature lacks studies that point to which emerging research directions need to be explored. This article proposes an agenda for future research based on the scarcity of literature in the field of event-oriented architecture. This agenda was derived from a literature review and a case study carried out, as well as from the authors' experience. Eight main topics were explored in this work: performance analysis, empirical studies, architectural stability, challenges to adopting, monitoring event streams, effects on software performance, broader challenges for adoption, and better monitoring of event-driven architecture. The findings reported help the researchers and developers in prioritizing the critical difficulties for uncovering the hidden potential of event-driven architecture. Finally, this article seeks to help researchers and professionals by proposing an agenda as a starting point for their research.
\end{abstract}


\maketitle

\section{Introduction}
\label{sec:intro}

The development of software systems currently takes place in increasingly unstable business environments, requiring high flexibility to support rapid system adaptations~\cite{oliveira2018brcode}. Typically, complex and volatile business rules, changes in used technologies, pressures for shorter development cycles and continuous delivery of functional modules~\cite{rubert2021effects} are some of the ever-present features of contemporary software development projects. For this reason, software development teams seek to use architectural styles, patterns, and software design principles to keep the stability of software system design under development or under maintenance~\cite{farias2014effects}. In addition, development teams seek to promote collaborative development, reduce development and cognitive effort~\cite{gonccales2019measuring,gonccales2021measuring}, improve effort estimates~\cite{carbonera2020software}, and minimize the effort invested to integrate and maintain critical software artifacts for the development of complex systems~\cite{farias2015evaluating,farias2013analyzing,bischoff2019integration}. In this sense, D’Avila~\cite{d2020effects} also outlines some findings about the effects of contextual information on the reduction of maintenance effort. Even worse, the lack of documentation for the design of software systems has been a factor that makes it even more difficult to maintain current software systems~\cite{junior2021survey,farias2018uml}. This absence of UML models in software projects, in part, is motivated by the difficulty of objectively evaluating the generated UML models~\cite{farias2020s} and keeping such UML models up to date with each other~\cite{de2019umlcollab} and with the current version of the source code.

Some works~\cite{stopford2018designing, Cao2021, Surantha2022, khriji2022design} point out that event-driven architecture promotes loose coupling --- essential for the modularization of application services --- but can increase design complexity and system understanding \cite{Ludger2002}. The software industry has adopted the use of events in architectures. For example, Spotify introduces streaming event delivery to batch to support its applications~\cite{Spotify}. Recent studies~\cite{laigner2020monolithic, Surantha2022, khriji2022design} point out the possible benefits of event-driven architecture. Laigner \textit{et al.}~\cite{laigner2020monolithic} report an empirical study in which the adoption of event-driven architecture improved the maintenance and fault isolation in a system that was refactored after years of maintenance, giving rise to a large and complex source code.

The literature on event-driven architecture advocates that designing applications strongly based on events favors the functionality modularization, as well as facilitating maintenance activities and service evolution of applications \cite{Cao2021}. In this sense, designing software adopting event-driven architecture may imply a more systematic way to promote a better modularization of modern software. Moreover, event-driven architecture has been widely adopted in the software industry, emerging as an alternative to modular development to support rapid adaptations of constantly evolving systems. However, little is known about the effects of event-driven architecture on performance, stability, and software monitoring, among others. Consequently, professionals end up adopting it without any empirical evidence about its impact. Even worse, the current literature lacks studies that point to which emerging research directions need to be explored. 

This article, therefore, aims to identify the challenges, implications, and future research directions regarding the use of event-driven architecture in the field of software development. For this, we carried out a scoping review in the literature, selecting works related to event-driven architecture for convenience, as there are few public studies on the areas covered, they adopt well-known research guidelines \cite{wohlin2012experimentation}. Moreover, a case study on the software modularity of event-driven architecture was designed and run. Based on these studies, we propose a research agenda that can be exploited to foster future research as well as initiatives in the software development industry. As a result, the article seeks to promote the study and adoption of event-driven architecture. In particular, researchers and professionals can benefit from this work. Researchers can benefit from this research by using it to build their research agendas, direct research efforts, and serve as a guide or starting point for students. Professionals can use this article as support to improve software development practices, drive architectural improvements, and promote software redesign.

This study is structured as follows: Section~\ref{sec:event-driven} introduces the main concepts for understanding the event-driven architecture; Section~\ref{sec:related-works} addresses related works, exploring the comparing them with the present one; Section~\ref{sec:methodology} describes the methodology for developing this study; Section~\ref{sec:future-directions} brings the proposed agenda; and, finally, Section~\ref{sec:conclusion} draws some conclusions and future works.

\section{Event-driven architecture}
\label{sec:event-driven}

\begin{figure}[!t]
\centerline
{\includegraphics[width=8cm]{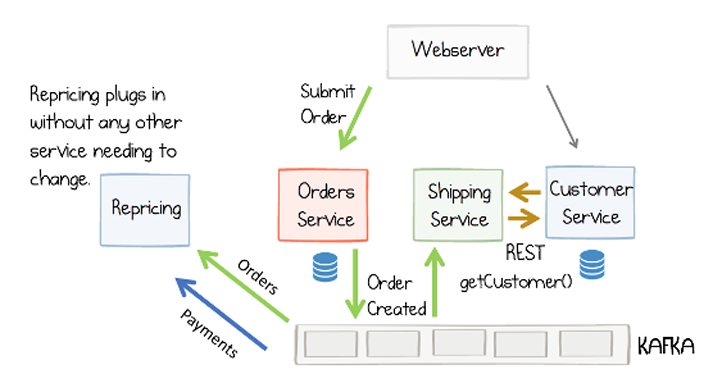}}
\caption{Flow of events in the ordering service (source \cite{stopford2018designing})}
\label{fig:eda}
\end{figure}

In event-driven architecture, software components publish data without knowing the other components or which can consume and react to the published data, promoting the separation of computation and event publishing from any subsequent processing~\cite{Ludger2002, khriji2022design}, as illustrated in Fig.~\ref{fig:eda}. Furthermore, the communication between the producer/consumer is asynchronous, and both are independent of each other \cite{Falatiuk2019}. Consequently, promoting loose coupling between components --- that is the reason event-driven architecture has become predominant in large-scale distributed applications \cite{Ludger2002}.

In addition, three-dimensional visual reconstruction using trains of events with very high temporal resolution, simulation of spiking neural networks and integration of multi agent systems are among other applications that EDA plays an important role. Such applications have some requirements in common, such as responsiveness, event asynchronicity, and heterogeneous data source \cite{schipor2019euphoria}.

The messaging system allows building loosely coupled services, as it moves the raw data to a highly coupled location (producer) and places it in a loosely coupled location (consumer) \cite{stopford2018designing}. Therefore, any operations that need to be performed on this data are not done on the producer, but on each consumer \cite{stopford2018designing}. That is, services can easily be added to the system in plug and play (pluggable) mode, where they connect to event streams and run when their criteria is met \cite{stopford2018designing, khriji2022design}. It does not only promote loose coupling, but also manage to store events and data, dispensing with the use of a database, keeping events ``close'' to the services \cite{stopford2018designing}. In addition, all events are stored in the order they arrived, allowing events to be played back in order. As a result, the performance of event-based applications is also better, ensuring stability and high performance for high data flow \cite{stopford2018designing}.

Its composition generally comprises components that detect events, listen for events, process the reaction to an event, and transmit events or messages between components. Event-driven architecture is extremely loosely coupled and highly distributed by design \cite{Cao2021}. On the other hand, decoupling between producers and consumers makes controlling the visibility of data more difficult. Although there are few studies addressing the disadvantages or difficulties with EDA, some points prove to be quite challenging.

\section{Related Work}
\label{sec:related-works}

This section discusses works that are close to the objectives of our article. The selection of related works was carried out following two steps: (1) search in digital repositories, such as Google Scholar for articles applying the search string ``event-driven architecture OR event-driven OR EDA''; and (2) filter of selected articles considering the alignment of such works with the objective of the work and the formulated research questions (Section~\ref{subsec:objective-research-questions}). We selected five articles from the literature for convenience, using the criterion of proximity to the topic explored in our research. Such works are analyzed (Section~\ref{subsec-works}) and then compared with the proposed work, aiming to identify research opportunities (Section~\ref{subsec:comparative-related-works}). 

\subsection{Analysis of Selected Studies}
\label{subsec-works}

\textbf{Overeem \textit{et al.} (2021) \cite{Overeem2021AnEC}.} The study analyzes 19 event-sourced systems to understand the reasons for using the event sourcing pattern. Its conceptualization is addressed for better understanding, as well as differentiating from similar architectures such as event-driven, although very similar, they present differences on event concepts. To understand the rationale for adopting event sourcing, it was based on interviews with 25 event sourcing engineers. Through the analyzed systems, three topics associated with event sourcing emerged, the use of Domain-driven Design (DDD) as software design, Command and Query Responsibility Segregation (CQRS) being a related architectural standard and microservice as a style. In each topic the reasons for application are discussed. In addition to the favorable points, five challenges faced by professionals are discussed: event system evolution, the steep learning curve, lack of available technology, rebuilding projections, and data privacy. Finally, from the insights acquired in the analyzed systems and the challenges found, five tactics and solutions were discovered that support professionals in the evolution of event sourcing systems.

\textbf{Petrov \textit{et al.} (2021) \cite{Petrov2021}.} It analyzes the notions of event processing, event processing methods, area context, and urgent issues of event processing methods. Furthermore, approaches to its solution are proposed, differences in event processing methods are exemplified, and the disadvantages of the methods are highlighted. Emphasizing the following problems: out-of-sequence event processing; occurrence of duplicates; collisions in event processing; fault-tolerant distributed architecture; multithreaded event processing; adaptive load balancing circuits; event processing application monitoring. Various solutions are discussed and tested using the test bench in order to assess the consequences. Some methods can lead to performance degradation.

\textbf{Schmidt \textit{et al.} (2008) \cite{Schmidt2008}.} It conducted a survey of the current state of the art in event-driven architecture, with a focus on event and action processing. Where is described the prerequisites of an entirely new conceptual model to describe the reactivity that is closest to the way people react to events: based on the ability to identify the context during which active behavior is relevant and the situations in which it is necessary. Challenges for event processing are addressed as a way to manage a very valuable knowledge asset - knowing how to react (make decisions) in event-driven situations. By distinguishing between a non-logical and a logic-based view when dealing with event-triggered reactivity.

\textbf{Griffin and Pesch (2007) \cite{Griffin2007}.} It presents a survey of service-oriented architecture and web services in telecommunications. Its have gone through several changes and technological evolution, arising from regulations and competition. The article describes the changes in detail and shows that the need to adopt service-oriented architecture (SOA) in telecommunications has become an important item on the agenda of operators. To make this possible, event-driven architecture (EDA) is covered in detail, as SOA and EDA complement each other and are necessary in a real SOA implementation. While SOA provides a request/response message exchange, EDA is capable of long-term asynchronous processing.

\textbf{Chung-Sheng (2005) \cite{Chung2005}.} It conducted a study on the evolution of event-driven applications and discusses their potential implications for system and middleware trends. Applications include: telecommunications services, trading system for financial services, logistics and asset management in manufacturing, digital fuel field for oil and gas production, telematics for automotive maintenance, and disease monitoring for healthcare. There is also the analysis of new middleware components and system architectures optimized for event processing and routing.

\subsection{Comparative Analysis and Research Opportunities}
\label{subsec:comparative-related-works}

\textbf{Comparison Criteria.} Seven Comparison Criteria (CC) were defined to identify the similarities and differences between the proposed work and the selected articles. This comparison seeks to help identify research opportunities using objective rather than subjective criteria. The criteria are described below:

\begin{itemize}
    \item \textbf{Agenda research (CC1):} understands studies to address future work or agenda research;
    \item \textbf{Event-driven architecture (CC2):} studies that address concepts or applied the event-driven architecture;
    \item \textbf{Empirical Study (CC3):} studies that performed experimental studies, especially through case studies, experiments, or observations for data collection in the field;
    \item \textbf{Performance analysis (CC4):} studies where performance analyzes were applied;
    \item \textbf{Stability analysis (CC5):} studies where software stability was measured;
    \item \textbf{Challenges to adopting (CC6):} studies that explore the challenges for the adoption of event-driven architecture;
    \item \textbf{Logs (CC7):} studies covering techniques for log analysis in event-driven architecture applications.
\end{itemize}

\textbf{Research opportunities.} Table~\ref{tab:comparative-table} presents the comparison of the selected studies, highlighting the similarities and differences between them. To sum up, the analyzed works do not present a purposeful research agenda which can serve as a starting point for future works. Therefore, this work seeks to fill this gap in the current literature.

\begin{table*}[!ht]
    \centering
    \begin{tabular}{|l|c|c|c|c|c|c|c|c}
        \hline
        \multicolumn{1}{|c}{\multirow{2}{*}{\textbf{Related Work}}}
        & \multicolumn{7}{c|}{\textbf{Comparison Criteria}} \\\cline{2-8}
         & \textbf{CC1} & \textbf{CC2} & \textbf{CC3} & \textbf{CC4} & \textbf{CC5} & \textbf{CC6} & \textbf{CC7} \\ \hline
        Proposed work & $\CIRCLE$ & $\CIRCLE$ & $\CIRCLE$ & $\CIRCLE$& $\CIRCLE$ & $\CIRCLE$ & $\CIRCLE$    \\ \hline
        
        Overeem \textit{et al.} (2021) \cite{Overeem2021AnEC} 
        & $\Circle$  & $\Circle$ & $\Circle$ & $\CIRCLE$ & $\Circle$ & $\CIRCLE$ & $\Circle$ \\ \hline
        
        Petrov \textit{et al.} (2021) \cite{Petrov2021}    
        & $\Circle$  & $\CIRCLE$ & $\CIRCLE$ & $\CIRCLE$ & $\Circle$ & $\CIRCLE$ & $\Circle$   \\ \hline
        
        Schmidt \textit{et al.} (2008) \cite{Schmidt2008}  
        & $\Circle$  & $\Circle$ & $\Circle$ & $\Circle$ & $\Circle$ & $\CIRCLE$ & $\Circle$   \\ \hline
        
        Griffin and Pesch (2007) \cite{Griffin2007}        
        & $\CIRCLE$  & $\CIRCLE$ & $\Circle$ & $\Circle$ & $\Circle$ & $\CIRCLE$ & $\Circle$   \\ \hline
        
        Chung-Sheng (2005) \cite{Chung2005}                
        & $\Circle$  & $\CIRCLE$ & $\Circle$ & $\CIRCLE$ & $\Circle$ & $\Circle$ & $\Circle$ \\ \hline
    \end{tabular}
    \resizebox{7cm}{!} {
    \begin{tabular}{cc}
        $\CIRCLE$ Attends & $\Circle$ Does not attend
    \end{tabular}}
    \caption{Comparative analysis of selected relate works}
    \label{tab:comparative-table}
\end{table*}

\section{Methodology}
\label{sec:methodology}

This section defines the methodology used to retrieve the state-of-the-art literature about research agenda regarding the subject of event-driven architecture in software engineering. From this selected literature, we derived the further challenges and implications of applying the event-driven architecture in industry. Section \ref{subsec:objective-research-questions} presents the objective and research questions of this work. Section \ref{subsec:search-strategy} presents the strategy to select the potential studies. Section \ref{subsec:study-selection} outlines the process of study selection. Section~\ref{subsec:case-study} briefly describes the exploratory case study run to identify the benefits of using event-driven architecture in terms of modularity.

\subsection{Objective and Research Questions}
\label{subsec:objective-research-questions}

This study has two main objectives: (1) to grasp a research agenda in relation to the event-driven architecture in software engineering; and (2) to list which implications that the software industry needs to overcome. For this, this study comprises in one research question: How to propose a research agenda containing challenges, implications, and future directions regarding event-driven architecture? 



To answer this question, a scoping review was conducted to select systematically the related literature.

\subsection{Search Strategy}
\label{subsec:search-strategy}

This section presents the strategy used to search studies in the current literature. The strategy consists of building a search string and defining a main search engine to conduct the search. 

\textbf{Search string.} The search string for related works was constructed using the main terms of event-driven architecture. The search string used was as follows:

\begin{tcolorbox}
``event-driven architecture OR event-driven OR EDA''
\end{tcolorbox}

In addition to works related to event-driven architecture, works referring to other fields of research are also returned. To address the topics on the agenda, specific research was carried out on each topic that generated other search strings in addition to the one mentioned above, aiming at the most relevant works regardless of the technologies involved.

\textbf{Search engine.} In this work, Google Scholar was the search engine used in the selection of works. This engine was selected because it encompasses works published in the main magazines and congresses related to computer science, which contain works related to event-driven architecture.

\subsection{Study Selection}
\label{subsec:study-selection}

The criterion for selecting the works, in short, was for convenience. Because, the area that studies event-driven architecture still has few works that address the themes explored in this study. On top of that, most of the jobs returned by the search engine were related to other fields of research. Therefore, it becomes impossible to apply the selection of works as some methodologies apply based on this return. Thus, the selection criterion was the title of the article and its abstract, enough points to understand if the article addresses the research objectives.

\subsection{Case Study}
\label{subsec:case-study}

As previously mentioned in Section~\ref{sec:intro}, event-driven architecture has been widely adopted in the software industry, thus emerging as an alternative to the development of enterprise applications based on REST architectural style, for example. We realize that little was known about the effects of event-driven architecture on software modularization while enterprise applications evolve. Consequently, practitioners ended up adopting it without any empirical evidence about its impacts on essential indicators, including separation of concerns, coupling, cohesion, complexity and size \cite{luanArvix}. We designed and run an exploratory case study to compare event-driven architecture and REST style in terms of modularity. A real-world application was developed using an event-driven architecture and REST through five evolution scenarios. In each scenario, a feature was added. The generated versions were compared using ten metrics. The initial results suggest that the event-driven architecture improved the separation of concerns, but was outperformed considering the metrics of coupling, cohesion, complexity and size. The findings are encouraging and can be seen as a first step in a more ambitious agenda to empirically evaluate the benefits of event-driven architecture against the REST style.

This exploratory case study explored in practice the possible benefits of using event-driven architecture, which served to build the proposed research agenda.

\section{Proposed Research Agenda}
\label{sec:future-directions}

This section presents the proposed research agenda, which was structured through eight perspectives, including performance analysis, conducting empirical studies (Section~\ref{subsec:performance-analysis}), stability analysis (Section~\ref{subsec:stability-analysis}), challenges to adopting the technology (Section~\ref{subsec:adoption}), and logging (Section~\ref{subsec:logs}).

\subsection{Performance Analysis}
\label{subsec:performance-analysis}


The main characteristic observed in a distributed system is your \textbf{process capacity}. The performance of a software indicates the degree to which the system fulfills its tasks, measured by the response time and the efficiency with which it achieves it ~\cite{Kounev2008}. While response time is the time it takes to respond to a request. Sites are often vulnerable to high traffic, usually on special dates. However, this is often associated with system scalability. Thus, software performance and scalability are essential to avoid system downtime due to overload, something common in today's systems. Certain characteristics can affect these features, e.g., message size, device type, and the complexity of the smart environment in terms of the number of producers and consumers \cite{schipor2019euphoria}.

For that, it is important to explore how the event-driven system should be configured to extract its \textbf{scalability} to the maximum~\cite{khriji2022design}. Scalability refers to the software's ability to deliver a continuous response time and throughput as the demand for the service it provides increases and resources are added~\cite{Kounev2008, khriji2022design}. It can be vertical or horizontal. Cloud environments allow the infrastructure team to easily configure as many machines as they need to meet demand. In addition to allowing you to optimize operating costs, once the environment is well configured and optimized. The literature~\cite{stopford2018designing} explores the capabilities in terms of configuration of event-driven platforms, such as Kafka. As the design and implementation of distributed systems is a challenge, ensuring that the software achieves the analysis objectives is crucial. Therefore, comparative studies on different platforms have become opportunities to explore their capabilities, difficulties and best approach. A key challenge is an initial assessment of performance-related factors, where potential bottlenecks in modeling, design, and implementation can be identified. This can reduce the cost of making any changes~\cite{Razib2015}. Along these lines, performance comparisons between event-driven architecture and traditional architectures such as REST are also interesting.

The event-driven architecture stands out for its high event processing capacity, handling continuous and \textbf{high-volume} data streams~\cite{Rahmani2021, khriji2022design}. In addition to being chosen for cases that demand scalability, it can be scalable in streaming events, scalable in the volume of historical data, scalable in the number of sources and data collectors, scalable in the number of processing elements, and scalable in terms of physical infrastructure \cite{Fournier2015, khriji2022design}. Moreover, it facilitates the interaction of heterogeneous devices in intelligent environments, where each one operates on its own operating system, communication protocol, a form of interaction, among others~\cite{schipor2019euphoria}.


\subsection{Empirical Studies}
\label{subsec:empyrical-studies}


Due to the differences pointed out between event-driven architecture and traditional architectures, naturally there must be \textbf{difficulties on EDA}. While in traditional software architectures, functions are nested, and it is easy to locate which functions call a given function. In event-driven architecture, events trigger services by each service criterion. Therefore, knowing which functions participate in a given flow becomes more challenging. Therefore, it is essential to know what can happen and how to deal with these difficulties, before the problem happens. The most dangerous time to deal with this is during a crisis, as taking actions without knowledge can make it worse.

Therefore, studies that address the \textbf{drawbacks and benefits} of event-driven architecture are essential for adoption. In Laigner \textit{et al.}~\cite{laigner2020monolithic}, a monolithic application was replaced, motivated by the difficulty in the maintenance process. The legacy BDS application was replaced by a current Big Data application, some benefits were perceived such as ease of maintenance and fault isolation. However, the complex data flow generated by the amount of microservices, as well as the myriad of technologies, have drawbacks.

The literature presents several case studies involving event-driven architecture. As a Big Data system (BDS) application for oil industry traffic monitoring~\cite{laigner2020monolithic}, Euphoria is a new software in event-driven architecture aimed at intelligent environments~\cite{schipor2019euphoria}, Digital Twin real-time data stream processing system~\cite{Alaasam2019}, e-archive document management system~\cite{Falatiuk2019}. In general, event-driven architecture is used to meet imposed needs, such as modularity, scalability, and asynchronicity for producing, processing, and transmitting messages and events~\cite{schipor2019euphoria, Falatiuk2019}.


\subsection{Stability analysis}
\label{subsec:stability-analysis}

Less change propagation in \textbf{maintenance} is desirable for any system. Since software lifecycle costs, it is estimated that between 40\% and 67\% are related to software maintenance ~\cite{Yau1980}. Therefore, software maintenance can be seen as one of the main attributes of software quality. A simple change in Service A can interrupt a Service B, simply because a function call has some parameter added or removed. Generating some unavailability in these services. In event-oriented architecture, the weak connection between services makes it more difficult to track this, since one event triggers another, when it meets the  conditions. Therefore, changes to one service can propagate to others, so documentation about these relationships between services can be helpful.

Once the maintenance represent a high cost in a software lifecycle, stable software is aimed at. For this, studies that explore the \textbf{stability} promoted by the event-driven architecture explain benefits in terms of lower costs. The first attribute affected when making a software modification is stability. If the software stability is low, the impact of any modification will be high. Therefore, maintenance will cost more and reliability can also suffer due to the introduction of possible new errors~\cite{Yau1980}. The stability of a module can be defined as the resistance of a change in one module to affect other modules~\cite{Yau1980}. Plus, there's logical and performance stability. Where logic measures the impact of a change in one module propagating to another, in logical terms. The performance measures the impact of changes considering the performance~\cite{Yau1980}. The literature points out that the highest costs involving software are related to the lack of average software measures. Such measures can be attributed to software quality, which is quite generic~\cite{Yau1980}. Studies in this area have contributed to define some software quality attributes such as correctness, flexibility, portability, efficiency, reliability, integrity, testability, and maintenance. Therefore, stability can be considered a very relevant metric for any software. Because, as studies show, it is directly linked to software maintenance. Comparisons between the stability promoted by event-driven architecture and traditional architectures will be able to show which tends to be less expensive.

This shows that the \textbf{software evolution} based on event-driven architecture has considerable challenges to explore. Current literature~\cite{stopford2018designing, Cao2021} points out that event-driven architecture promotes loose coupling --- essential for the modularization of software services --- but can increase design complexity and system understanding~\cite{Ludger2002}. Such modularization aims to isolate the software modules so that changes that occur in one module do not affect the others, so the software would present a better stability. This modularity can be useful in scenarios where services have been added/removed in system software. Since modularity prevents maintenance propagation. This can be measured through metrics collected between versions of a software system's evolution.


\subsection{Challenges to Adopting}
\label{subsec:adoption}

Systems that want to take advantage of the benefits of event-driven architecture will have some challenges. It would be interesting to \textbf{trace the types of systems} that benefit from event-driven architecture. A good architecture helps the system meet key requirements in areas such as performance, reliability, portability, scalability, and interoperability. Software architecture plays a key role as a bridge between requirements and implementation~\cite{Garlan2000}. By providing an abstract description of a system, the architecture exposes certain properties while hiding others. Ideally, this representation provides an intellectually traceable guide to the entire system, which allows designers to reason about the system's ability to satisfy certain requirements and to suggest a design for the construction and composition of the system~\cite{Garlan2000}. Sometimes when choosing the best architecture for a system, mistakes can be made when choosing the least suitable one, either because it is a trend or lack of knowledge about its strengths and weaknesses. In Overeem \textit{et al.} (2021) \cite{Overeem2021AnEC} the cause for the chosen event source for the systems under study was explored. Some respondents responded that it was because of the architectural trend and/or curiosity.

Another challenge is its \textbf{learning curve,} as event-driven architecture has considerable differences compared to traditional architectures. Building or rewriting systems will require teams composed of architects and a developer capable of handling event-driven architecture are needed to develop a system. However, most IT professionals deal with traditional architecture, and moving to event-driven architecture can be expensive. Garlan~\cite{Garlan2000} points out that software architecture can be useful in at least six aspects of the system: (i) \textit{understanding}: simplifies our ability to understand systems, by the abstraction presented in which the high-level design of a system can be easily understood; (ii) \textit{reuse}: it is divided into several levels, such as reuse of components, as well as structures in which they can be integrated; (iii) \textit{construction}: provides a diagram indicating the main components and dependencies between them; (iv) \textit{evolution}: expose the dimensions along which the system is expected to evolve, as well as assist in maintenance, exposing the unfolding of changes and, thus, more accurately estimating the costs involved; (v) \textit{analysis}: provides opportunities for analysis, including consistency checking, compliance with constraints imposed by an architectural style, compliance with quality attributes, domain analysis, and dependency for architectures built in specific styles; (vi) \textit{management}: successful projects see software architecture as an important milestone in the development process, as it makes requirements, implementation strategies and potential risks much clearer.


\subsection{Logs}
\label{subsec:logs}

Capturing logs from event streams is one of the challenges that could be researched and deepened. Despite efforts to deliver working software, the size and complexity of the software, combined with the real time and budget for development, make it increasingly difficult to deliver fully working software~\cite{Ding2011}. Software systems inevitably have flaws, such as bugs triggered by some combination of errors, environmental issues, or administrative errors. However, it is not always easy to identify such flaws, which can take time. They are quite worrisome for two reasons: generally they must be resolved quickly, as it is in production, so some part of the software must be inaccessible; second, the difficulty in analyzing the failure, commonly due to lack of data, making reproduction and assertiveness in the treatment impossible. A tool widely used by developers is the System Logs, they make it possible to track and record the behavior at run-time of the software. They are usually used for monitoring, fault diagnosis, performance analysis, test analysis, security and legal compliance and business analysis~\cite{Boyuan2021}. 

Logging consists of two phases: (i) instrumental logging and (ii) log management. Instrumental logging consists of code inserted by developers to record information at run-time. Log management is concerned with analyzing the collected logs to generate relevant information about the behavior at run-time~\cite{Boyuan2021}. Therefore, extracting information from logs in log management depends on the quality of the log code produced in the previous phase. The low quality of the log can imply in the diagnosis of problems, high effort in maintenance, low performance, or even software crashes~\cite{Boyuan2021}. Therefore, it is important to record the system logs, mainly to identify failures, which can make part of the system unavailable, as well as their analysis to extract relevant data about the system's functioning. To mitigate possible failures, improve performance and even help with maintenance. Tools commonly applied in other technologies can be used to capture logs. However, high event traffic in an event-driven system will likely generate a lot of data, so feasibility studies or better techniques are essential.

As a means of protection to downtime, the service contracting party has Service Level Agreements (SLAs) which are a service contract in which the service level between the customer and supplier is defined. SaaS providers are responsible for ensuring that applications are available 24 hours a day, 7 days a week (24/7), ideally with no downtime. Downtime tolerance in SaaS apps can be less than traditional web apps. The effects of downtime on SaaS applications match the effects of downtime on large-scale e-commerce applications~\cite{Banerjee2010}. This can lead to significant financial losses. For that, availability measures are used to define uptime and downtime, such as response time for requests, as well as possible penalties in case of downtime under the provider.

In addition, \textbf{distributed log composition} is also worth studying. As several differences can be observed in relation to traditional architectures, mainly in the execution, which occurs individually in each service as opposed to a stack trace as in traditional architectures. Events are triggered by others, there is no traditional stack, therefore, how to identify that a service is not being triggered. Tools like Automated Log Abstraction Techniques (ALAT) can help. However, there is a gap between academia and industry as engineers do not know the best ALAT, due to lack of time and resources to search the literature, and due to lack of studies that describe the best ALAT~\cite{ElMasri2020}. Therefore, logs contain abundant data that can help engineers understand the runtime properties of a system. However, large and complex systems produce abundant data to analyze. For this, ALAT can help reduce the data to be processed through its record abstraction algorithms~\cite{ElMasri2020}.

\subsection{Effects on Software Performance}

The adoption of event-driven architecture has increased in recent years as an alternative to modular development. However, the impact of event-driven architecture on software performance remains unknown. The challenge for researchers is to investigate the impact of event-driven architecture on various aspects of software performance, such as CPU usage, memory usage, response time, throughput, and resource usage.

One challenge for researchers is determining the appropriate metrics to measure software performance accurately. For example, response time is often used as a performance metric for traditional modular systems, but it may not be the most appropriate metric for event-driven systems. Event-driven systems rely on asynchronous message passing, which can impact response time measurements. Thus, researchers need to consider new performance metrics that are specific to event-driven architecture to accurately measure software performance.

Another challenge is to design experiments that can simulate real-world scenarios accurately. Event-driven systems are often used in highly distributed environments, which can make it challenging to replicate realistic workloads in a controlled environment. Researchers need to design experiments that can accurately simulate real-world scenarios and workloads to measure the impact of event-driven architecture on software performance.

Finally, researchers need to consider the impact of the architecture on software performance over time. Event-driven systems are designed to support rapid adaptations of constantly evolving systems, which can impact performance over time. Researchers need to conduct longitudinal studies that can measure the impact of event-driven architecture on software performance over an extended period.

\subsection{Wider Challenges for Adoption}

While event-driven architecture has been adopted widely, professionals often adopt it without empirical evidence of its impact. The challenge for researchers is to explore the challenges for software adoption of event-driven architecture, including factors such as training requirements, development tooling, and team composition.

One challenge for researchers is to identify the specific training requirements necessary to effectively adopt event-driven architecture. Event-driven architecture requires a different mindset than traditional modular development. Thus, researchers need to identify the specific training requirements necessary for developers and operators to adopt event-driven architecture successfully. Training programs may need to focus on areas such as asynchronous programming, message passing, and distributed systems.

Another challenge is to identify the development tooling required to support event-driven architecture effectively. Traditional development tools may not be well-suited for event-driven systems. Thus, researchers need to identify and develop new tools that can support event-driven architecture, such as tools for message routing, data processing, and monitoring.

Finally, researchers need to consider the impact of team composition on software adoption. Event-driven systems require collaboration between developers, operators, and other stakeholders, which can be challenging in large organizations. Researchers need to explore the impact of team composition on the adoption of event-driven architecture and identify strategies for effective collaboration.

\subsection{Better Monitoring Techniques}

The lack of research on software monitoring for event-driven architecture is a significant challenge. The challenge for researchers is to develop better monitoring techniques that can help developers and operators to identify issues with event-driven systems in real-time, including techniques for monitoring system state, data flow, and message passing.

One challenge is to develop monitoring techniques that can handle the high message rates and data volumes associated with event-driven systems. Traditional monitoring techniques may not be well-suited for event-driven systems, which can generate large amounts of data quickly. Researchers need to develop new techniques that can handle the high message rates and data volumes associated with event-driven systems.

Another challenge is to identify the appropriate metrics for monitoring event-driven systems. Event-driven systems rely on asynchronous message passing, which can make it challenging to identify the source of issues when they arise. Researchers need to identify the appropriate metrics for monitoring event-driven systems, such as message queue lengths, message processing times, and error rates.

Finally, researchers need to consider the challenges of monitoring event-driven systems in distributed environments. Event-driven systems are often distributed across multiple nodes, making it challenging to monitor the entire system from a single point. Researchers need to develop techniques for distributed monitoring that can collect data from all nodes in the system and provide a unified view of the system state.

In addition, researchers need to consider the trade-offs between monitoring overhead and the accuracy of monitoring data. Event-driven systems are designed to be highly scalable, which means that monitoring overhead can impact system performance. Researchers need to develop monitoring techniques that strike a balance between overhead and accuracy, such as sampling techniques that can provide accurate data while minimizing monitoring overhead.

Overall, developing better monitoring techniques for event-driven architecture is crucial for ensuring the stability and reliability of these systems. The development of effective monitoring techniques will help developers and operators to identify issues in real-time and respond to them quickly, improving system uptime and reducing downtime.

\section{Conclusion}
\label{sec:conclusion}

Event-driven architecture has the potential architecture for the development of distributed systems, with promising gains in modularization, scalability and concurrency. Some studies show that it has been shown to be effective for various applications due to its characteristics. However, it has few empirical studies about performance, stability, log and challenges to adopting. Studies on these topics can accelerate the adoption by the industry, who has not certainly about the consequences. Therefore, the objective of this work was to provide future directions to researches and practitioners about the use of event-driven architecture on systems.

Overall, the agenda shows that event-driven architecture has many questions to be studied and presented to better understand the consequences of being adopted in software engineering. The challenges also reinforce when the architecture can represent a considerable part of the costs in the software lifecycle. Therefore, if event-driven architecture has the potential to decouple services, it could facilitate the maintenance process, consequently lowering the cost. We hope that this work will serve as a starting point for future research.

\section{Acknowledgment}

This work was partially supported by the Conselho Nacional de Desenvolvimento Cient\'{i}fico e Tecnol\'{o}gico (CNPq) under Grant 313285/2018-7.

\bibliographystyle{ACM-Reference-Format}
\bibliography{main}

\end{document}